\documentclass[a4paper,twoside]{article}

\usepackage{booktabs}
\usepackage{graphicx}
\usepackage[hidelinks]{hyperref}
\usepackage{epsfig}
\usepackage{subcaption}
\usepackage{calc}
\usepackage{amssymb}
\usepackage{amstext}
\usepackage{amsmath}
\usepackage{amsthm}
\usepackage{multicol}
\usepackage{pslatex}
\usepackage{apalike}
\usepackage{algorithm2e}
\usepackage[bottom]{footmisc}
\usepackage{framed}
\usepackage{xcolor}
\usepackage{SCITEPRESS}

\newcommand{\eg}{e.g.,\xspace}

\newcommand{\ie}{i.e.,\xspace}

\newcommand{\content}[1]{#1}

\newcommand{\point}[1]{\par\smallskip\noindent\textbf{#1}}

\newcommand{\toolname}{COBRA\xspace}
\newcommand{\fulltoolname}{Content-Agnostic Offline Bulk Registration Analyzer\xspace}

\newcommand{\vt}{VirusTotal\xspace}

\newcounter{finding}
\renewcommand{\thefinding}{\arabic{finding}}

\newcommand{\finding}[1]{
  \refstepcounter{finding}
  \begin{leftbar}
  \noindent\textbf{Finding \thefinding:} #1  
  \end{leftbar}
}

\begin{document}

\title{
\vspace{-1.8cm}
\begin{center}
\small
\textit{Published in the Proceedings of the 23rd International Conference on Security and Cryptography - Volume 1: SECRYPT, SciTePress, pages 37-48. DOI: \url{https://doi.org/10.5220/0015000800004103}}
\end{center}
\vspace{0.5cm}
\toolname: A Content-Agnostic Framework for Zero-Day Detection of Suspicious Domains}

\author{\authorname{Alexandros Fourtounis\sup{1,2}, Emmanouil Papadogiannakis\sup{1,2}, Panagiotis Papadopoulos\sup{1}, \\ Nicolas Kourtellis\sup{3} and Evangelos Markatos\sup{1,2}}
\affiliation{\sup{1}Foundation for Research and Technology - Hellas (FORTH), Heraklion, Greece}
\affiliation{\sup{2}University of Crete, Heraklion, Greece}
\affiliation{\sup{3}Keysight AI Labs, Barcelona, Spain}
}


\abstract{
The use of malicious domains is central to cyberattacks such as phishing, malware distribution, impersonation, and fraudulent transactions. 
Because domains are inexpensive to register and easy to deploy at scale, they remain one of the most common and damaging tools used in cybercrime across industries.
Proactive detection is essential to reducing this window of vulnerability and preventing harm to users.
In this work, we propose \toolname: a content-agnostic, registration-time detection framework for identifying and analyzing suspicious domains from day zero. 
Our approach does not rely on any content-based features, allowing us to classify a domain even before it is populated with content.
We analyze the names of newly registered domains and employ a clustering technique to group them based on lexical and structural similarity.
We evaluate our methodology using real-world data consisting of 1.5M newly created domains, demonstrating that \toolname detects suspicious domains with a precision of 98.5\%, identifying more than 47K distinct newly registered suspicious domains.
Furthermore, our results show that domain-name clustering enables accurate early detection, allowing us to identify 80\% of suspicious or malicious domains earlier than one of the most widely used threat-intelligence services, which in some cases may require up to 7 days.
}

\onecolumn \maketitle \normalsize \setcounter{footnote}{0} \vfill

\section{\uppercase{Introduction}}
\label{sec:introduction}

Domains are core infrastructure in cybercrime, with United States losses reaching \$16 billion in 2024~\cite{fbiCybercrime}.
Their low cost and scalability enable phishing, malware distribution, command-and-control (C2) communications, and scams by linking attackers, victims, and malicious services.
These activities impose significant direct and indirect costs, including fraud, breaches, ransomware (projected at \$57 billion annually by 2025~\cite{ransomwareCosts}), downtime, and reputational harm across individuals, organizations, and economies.

Research on malicious domain classification often focuses on lexical analysis and machine learning techniques.
Features such as domain length, n-grams, and entropy detect anomalous or algorithmically generated domains~\cite{zhao2019malicious,selvi2019detection}, while learning models capture more complex or semantic patterns~\cite{woodbridge2016predicting}.
However, only a few studies focus on detection at registration time, aiming to discover malicious or suspicious domains even before they become active~\cite{ccolhak2024securereg}.
Despite advances, the problem is yet to be solved because existing methods depend on historical data and overlook evolving naming strategies, which adversaries adapt to evade detection~\cite{liu2021cleter}.

Beyond academia, several commercial services classify suspicious domains, including 
Spamhaus, which maintains DNS blocklists for spam, phishing, and malware, 
Google Safe Browsing, which detects phishing and malware across the Web,
AbuseIPDB, which provides community-driven abuse tracking, and
Cisco Talos Intelligence, which offers enterprise domain reputation and threat intelligence.
Among the most widely used is \vt, which aggregates results from dozens of antivirus engines and reputation sources and is frequently cited in prior academic work (\eg~\cite{choo2023large,salem2021maat}).

Despite widespread deployment of detection systems, identifying malicious domains remains difficult.
Attackers can rapidly register domains at low cost, exploiting delays between registration and detection.
Our focus is on threat actors that rely on scale and automation.
For example, botnet operators use Domain Generation Algorithms (DGAs) to ensure secure communication between C2 channels. 
Additionally, adversaries often leverage APIs provided by registrars, to quickly deploy low-cost disposable domains.
Although blocking known malicious domains can disrupt attacks, newly registered or previously unseen domains often evade defenses until harm occurs.
The inability to reliably identify malicious domains early enables user interaction and malware communication with attacker-controlled infrastructure, leading to significant financial and operational losses. 
Addressing this gap requires more effective methods for the early detection and classification of malicious domains.

To address the limitations of existing approaches, we introduce \toolname (\fulltoolname), a novel framework for registration-time detection of malicious domains.
Unlike systems that rely on post-registration signals such as hosted content, DNS activity, or user reports, \toolname operates entirely offline at the moment of domain registration, enabling detection before domains are activated or abused.
\toolname is content-agnostic and designed to identify automatically generated and bulk-registered domains, a common characteristic of large-scale malicious campaigns.
Through extensive evaluation, we demonstrate that \toolname achieves a precision of 98.5\% while significantly reducing detection latency compared to state-of-the-art platforms such as \vt.
By eliminating the delay inherent in behavior-based detection, \toolname effectively closes a critical window of vulnerability (lasting up to 7 days in existing systems) and enables proactive disruption of malicious campaigns.

Overall, the contributions of this work are:
\begin{enumerate}

\item We propose \toolname, a novel content-agnostic framework for identifying suspicious domains at registration time, enabling proactive detection before domains are activated or abused.
Our analysis shows that, while \vt also employs a content-agnostic methodology, \toolname's advantage stems from its registration-time design, rather than differences in detection strategy.

\item We demonstrate that, by detecting suspicious domains at registration time, \toolname closes a critical window of vulnerability that can persist for up to 7 days when relying on \vt.
We further identify a consistent 20\% of suspicious domains that are never flagged by existing security engines, highlighting the limitations of current state-of-the-art approaches.

\item Through deployment on real-world registration data, we identify nearly 45K distinct suspicious or malicious domains, which we validate using \vt.
We make the resulting dataset publicly available to support reproducibility and future research~\cite{openSource}.

\item We show that \toolname remains robust across diverse website categories and detects suspicious domains missed by other security engines.

\item Finally, we analyze the malicious deployment strategies and find disproportionate use of specific registrars, such as GoDaddy, while frequently rotating across multiple registrars to evade detection. We also uncover extreme bulk registration, with a peak rate of 83 domains per second and an average rate exceeding one domain per second.
\end{enumerate}
\section{\uppercase{Suspicious Domains Detection}}
\label{sec:methodology}

To identify suspicious domains, we propose a novel algorithm that analyzes lexical similarities among domain names using a string distance function.
The goal of our method is to uncover patterns of illicit bulk registrations.
Our methodology relies on adversaries that register large batches of domain names within a short period of time.
We examine structural similarities in domain names and isolate clusters of similar domains that are automatically produced.
Our methodology relies solely on the newly registered domain names without the need for registration metadata or registrar behaviors to distinguish between benign and malicious bulk registrations.
The proposed methodology is proactive, that is, it identifies suspicious domains that will be used in malicious campaigns in the future.

Our methodology is applied on a set of domains registered on a given day. 
For each domain in the set, we compute pairwise string distances against all other domains, capturing similarity based on character-level substitutions.
Let two domains $d_1, d_2$, where $s_1$, $s_2$ are their respective second-level domains (SLDs) and $t_1$, $t_2$ are the top-level domains (TLDs), respectively.
We define the distance function $D$ over two domains $d_1$ and $d_2$ as:

\begin{equation}
     D(d_1,d_2) =
     \begin{cases}
         \sum_{i=0}^{n-1}\delta(s_{1,i}, s_{2,i}) & \text{if } t_1 = t_2 \\
         \infty & \text{if } t_1 \neq t_2
     \end{cases}
\end{equation}

The proposed distance function incorporates a strict TLD matching. 
If the two domains do not share the same TLD, their distance is defined as infinite.
The penalty is applied to ensure that domains with different TLDs will end up with an extensive distance.
Our decision to assign infinite distance across different TLDs is intentional and reflects a structural assumption in the threat model.
In practice, TLD boundaries correspond to different registration authorities, pricing models, policies, and reputation.
Consequently, domains under different TLDs are operationally and administratively independent, even if their second-level strings match.
In cases where domains share the same TLD, the distance is computed as the sum of character differences over the remaining prefixes of the two domains.
We use a character-level cost function $\delta(c_1,c_2)$ that measures the dissimilarity between individual characters $c_1$ and $c_2$.

\begin{equation}
\delta(c_1, c_2) =
    \begin{cases} 
        0 & \text{if $c_1, c_2 \in \text{Digits}$,} \\
        0 & \text{if $c_1, c_2 \in \text{Symbols}$,} \\
        0 & \text{if $c_1, c_2 \in \text{Letters and } c_1 = c_2$,} \\
        1 & \text{if $c_1, c_2 \in \text{Letters and } c_1 \neq c_2$,} \\
        1 & otherwise
\end{cases}
\end{equation}

The intuition behind the delta function is that it assigns a distance of 0 when the characters are considered equivalent, and 1 when they differ significantly.
All alphabetical characters are compared in a case-insensitive manner, while all digits and symbols characters are mutually equivalent.
That is, their distance is 0.
We intentionally disregard differences in numeric tokens and certain symbols, as these are frequently used as interchangeable placeholders in bulk domain registrations.
A distance of 1 is returned only when two characters belong to different classes, or when two letters are different.
For example, the distance between \emph{abc.com} and \emph{def.com} is 3 because all characters differ, whereas \emph{domain-12.com} and \emph{domain-23.com} have distance 0.
A small penalty is applied when a character differs in type (letter vs. digit or symbol), and string-length differences increase the distance for each missing character.
This function design focuses on structural similarity rather than exact character matching.
It does not capture every difference between two domain names, but instead groups many characters into equivalent classes.
Minor differences, such as character case, digits, or punctuation, are treated as noise and ignored.

To establish the similarity among multiple domain names, all possible domain name pairs are compared using our distance function $D$.
Resulting values are then aggregated into a distance matrix that encodes the lexical proximity of every domain pair.
We only consider domains whose pairwise distance is 0, focusing only on structural equivalence and ignoring small differences.
A distance of 0 indicates that domains have matching letters and only their digits or punctuation might differ.
A positive distance indicates that they are not equivalent, while an infinite distance suggests that the domains are not comparable.
Although alternative threshold values may suit specific applications, a study of different threshold strategies is beyond the scope of this work.
Domains with distance 0 are assigned to the same cluster, while domains not belonging to any cluster are treated as noise and excluded from further analysis.

We consider that all domains registered on exactly the same day that are placed inside a cluster (\ie highly similar domain names) can be treated as suspicious or, in some cases, malicious.
This approach aims to detect automated domain generation and bulk registration commonly used in malicious campaigns. 
Structurally indistinguishable domains found in the same cluster and with the same creation date, suggest automated domain generation rather than independent and legitimate domain registration.
\content{
Our approach does not rely on content.
This is a deliberate choice as content analysis requires a domain to be active, which inherently creates a window of vulnerability that can harm users.
In contrast, \toolname does not require a domain to be active or serving content. Thus, our approach is able to detect domains on day zero.}
\section{\uppercase{Threat Validation}}
\label{sec:validation}

\subsection{Data Collection}
\label{sec:data-collection}

Malicious operators often register large batches of domains within short time windows to support rotational or disposable infrastructure.
It is unlikely that multiple unrelated benign entities independently select domains from the same structural cluster (under our distance function) and register them on the same day.
To capture this behavior, we track each domain's registration date and collect newly registered domains daily.
We make use of a commercial provider\footnote{\url{https://domains-monitor.com}} that supplies us with daily updates of zone files.
We have no affiliation with the specific provider and only use this service to extract newly registered domains.

We randomly select 6 unpopular TLDs with accessible zone files: \emph{.bond}, \emph{.life}, \emph{.sbs}, \emph{.space}, \emph{.store} and \emph{.top}.
We deliberately select unpopular TLDs, as attackers may favor them for discounted or promotional registration fees, enabling bulk domain registration at minimal cost~\cite{lim2025registration}.
Additionally, unpopular TLDs have low registration density, meaning that algorithmically generated names are more likely available with no collisions.
For instance, as of Dec 4, 2025, the \emph{.com} TLD contains 158M distinct domain names, while \emph{.sbs} only 1M, allowing adversaries to generate similar and recognizable domains.
Compounding the issue, some registrars have minimal verification, letting malicious 
actors register multiple domains in bulk~\cite{lim2025registration}.

Traditional detection often observes suspicious behavior only after a domain is active and potentially used for phishing, malware, C2, or fraud.
In contrast, registration-time detection allows proactive intervention before the malicious infrastructure becomes operational.
Towards that, we build \toolname (\fulltoolname), a content-agnostic framework that detects suspicious domains at registration time.
\toolname fetches the zone files of the selected TLDs and extracts newly registered domains on a daily basis.
Utilizing the distance function of Section~\ref{sec:methodology}, it uncovers new suspicious domains while they are is still inactive and prevents harm rather than responding to it.

\subsection{Detection Validation}
\label{sec:detection-validation}

We start by forming the hypothesis that all domains which have been clustered together by our algorithm are suspicious.
To verify and compute the accuracy of our hypothesis, we deploy our detection methodology on a real-world scenario.
\toolname provides an automated pipeline that obtains newly registered domains and clusters suspicious ones on a daily basis.

To evaluate the correctness of our decision that a set of domains is suspicious, we create a ground truth dataset based on information provided by \vt, one of the most popular and widely used threat intelligence services.
\vt combines results from dozens of antivirus engines, URL scanners, and threat-intelligence providers.
We consider a domain suspicious if it has a flag score greater than 0, or if at least one of the security engines marks it as not clean.

While cross-validation with multiple independent providers would be desirable, this is challenging since commercial services often impose strict usage limitations and pricing models that make large-scale evaluation infeasible.
Moreover, the labeling methodologies of different engines differ significantly.
In preliminary experiments, we observe discrepancies across services, including cases where well-established domains (\eg \emph{google.de}) are flagged as malicious, or where entire TLDs (\eg \emph{.zip}) were categorized as suspicious without any domain differentiations.
We therefore rely on \vt as an aggregation platform which is not tied to a single security vendor's detection logic, thus providing a more reliable signal.
\vt was additionally selected due to its widespread use in existing academic work (\eg~\cite{alsabah2022content,deniz2025mantis}).
However, our methodology is innately feed agnostic and does not rely on a specific threat intelligence to function.
It can be easily extended to use alternative threat intelligence sources, other than \vt.

To evaluate the performance of \toolname, we deploy it on real-world data 
for a duration of ten consecutive days, starting September 10, 2025.
Each day, we extract all newly registered domains for 6 TLDs and deploy our detection methodology to identify clusters of suspicious domains that are bulk registered.
We consider all domains clustered by \toolname as suspicious.
To verify our assumption and measure the correctness of our methodology, we cross reference our findings against \vt.
If \vt flags a domain, we immediately consider it as suspicious and attribute it as a correct classification by \toolname.
Otherwise, we perform a manual security investigation to determine whether a domain is indeed malicious but (temporarily) undetected by security engines.
We deliberately restrict our manual analysis to features that can be objectively associated with suspicious or malicious behavior, rather than relying on the subjective judgment or personal opinions of the investigators.
Consequently, for domains that are active and respond to requests, we perform a multistep analysis and collect any potential HTTP or JavaScript redirections to other domains, any served smartphone applications (\ie APKs) that can be downloaded from the website, as well as all URLs found within the page's HTML code.
All these website traces are then fed into \vt for a more granular security evaluation.
If any of the collected traces is flagged by \vt, we determine that the initial domain we visited is also suspicious.
This decision is solely based on the output of \vt and does not rely on judgments of the investigators.
This multi-step analysis uncovers hidden maliciousness, as some domains are indirectly suspicious through connections to other malicious resources (\ie content they host, link to, or distribute).

In addition to establishing the ground truth, it is necessary to evaluate domains on a daily basis.
A domain that is classified as benign on a given day may be repurposed for malicious activity on subsequent days, and thus ongoing assessment is required to capture such changes.
We set 10,000 as the maximum number of clustered domains for this experiment, to ensure that we can timely collect all needed information each day.
We stop adding new domains daily after we reach the set threshold, which is met after only three days of collecting newly registered domains.
We provide a summary of suspicious domain detection results per TLD in Table~\ref{tab:methodology-validation}, showing the number of domains identified by our approach, the corresponding counts reported by \vt, and the relative detection rate.
We discover that more than 68\% of detected domains are active at some point throughout the experiment, meaning that potential victims could be targeted until the domains are flagged and taken down.

\begin{table}[t]
    \centering
    \scriptsize
    \begin{tabular}{lrrr}
    \toprule
        {\textbf{TLD}} & {\textbf{Detected Domains}} & {\textbf{Determined Suspicious}} & {\textbf{Rate}}\\
        & & {\textbf{by VirusTotal}} \\
    \midrule
        \emph{.bond}  & 2329 & 2329 & 100.0\% \\
        \emph{.life}  &   41 &   32 & 78.05\% \\
        \emph{.sbs}   &  451 &  451 & 100.0\% \\        
        \emph{.space} &   24 &   22 & 91.67\% \\
        \emph{.store} &  178 &  106 & 59.55\% \\
        \emph{.top}   & 7822 & 7748 & 99.05\% \\
    \bottomrule
    \end{tabular}
\caption{Comparison of the number of domains detected as suspicious by our method versus \vt along with the detection rate relative to \vt.}
\label{tab:methodology-validation}
\end{table}

We find that our detection methodology achieves an overall precision of 98.52\%.
That is, only about 1.48\% of the domains flagged as suspicious by \toolname are false positives, demonstrating that our method is highly reliable in the domains it identifies as suspicious.
The proposed method is designed to prioritize precision over recall.
Our goal is to ensure that domains identified as suspicious are highly likely to be truly suspicious, even if this approach does not detect all suspicious domains.
A lower recall is acceptable because the primary goal is to make confident, reliable classifications.
Missing a subset of suspicious domains is less critical in this work than falsely labeling benign domains as malicious.
Finally, estimating false negatives requires ground-truth knowledge of all malicious domains registered on a given day, which is not practically obtainable due to rate limits and the absence of daily labeling.
Our evaluation prioritizes precision, focusing on being confident that the domains flagged by our system are suspicious, rather than detecting all malicious domains.

We select low-popularity TLDs for evaluation because reports suggest that bad actors often prefer such TLDs due to low registration costs, weaker oversight, and reduced scrutiny compared to highly regulated popular TLDs.
In preliminary experiments on popular TLDs, \toolname exhibits similar performance trends but with fewer detections due to the lower rate of malicious domains.
For example, out of 8,052 \emph{.net} domains registered in one day, \toolname flagged 210 as malicious, achieving 97.6\% precision against \vt.
For \emph{.org}, 52 of 9,256 domains were flagged, with 41 confirmed by \vt (78.8\% precision).
These results demonstrate that \toolname maintains high precision even on popular TLDs, though absolute detections are limited by the low prevalence of abuse.
We therefore focus on less popular TLDs to ensure sufficient detections for robust analysis.

\subsection{Bulk Registration}

To further analyze the nature of clusters identified by \toolname, we examine the registration times of domains within each cluster.
Suspicious domains are often registered in large batches with patterns that cannot correspond to manual actions or benign registration behavior.
Using the Registration Data Access Protocol (RDAP)~\cite{rdap}, we successfully collect registration information for $\sim$39K distinct suspicious domains identified by \toolname over a period of multiple months (see Section~\ref{sec:real-world-deployment}).
We observe that domains belonging to the same cluster exhibit highly concentrated registration timestamps, often occurring within very short time periods.
The median cluster contains domains that are registered within a window of just five minutes.
80\% of clusters consist of domains registered within one hour, while notably, 46.1\% of clusters contain domains registered within a single minute.
Tightly grouped registration times indicate automated bulk registration, supporting the idea that domains in each cluster are generated programmatically.

\begin{figure*}[t]
    \begin{minipage}[t]{0.32\textwidth}
        \centering
        \includegraphics[width=1\columnwidth]{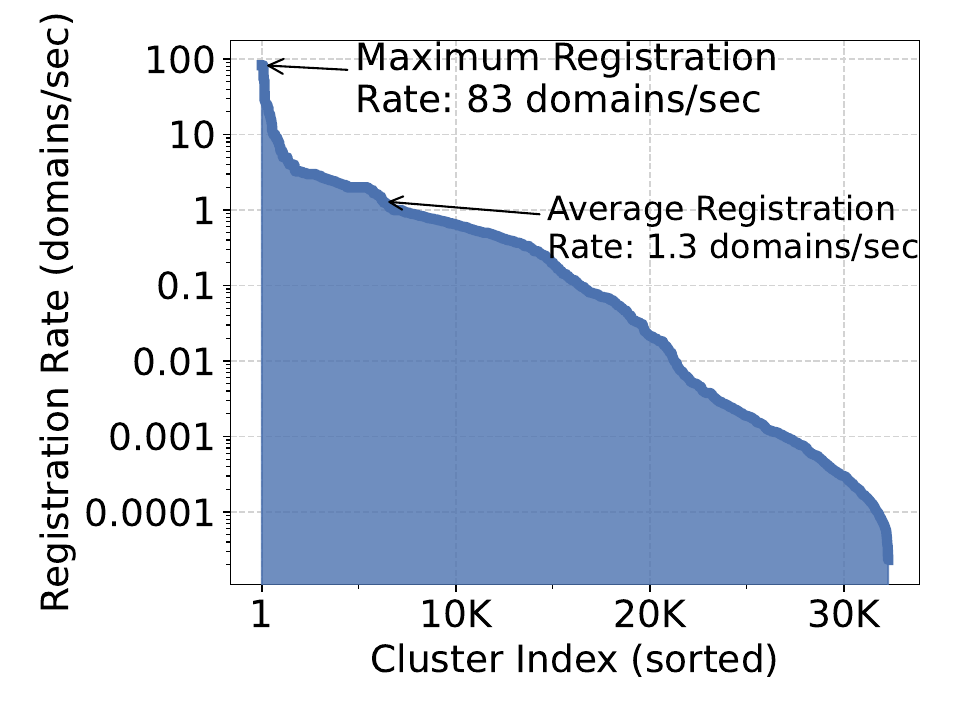}
        \caption{Distribution of registration rate across clusters, weighted by the number of websites per cluster ($y$-axis in log scale). Registration rate measures the intensity of automated bulk registration, with clusters averaging more than one domain registered per second.}
        \label{fig:registration_times}
    \end{minipage}
    \hfill
    \begin{minipage}[t]{0.32\textwidth}
        \centering
        \includegraphics[width=1\columnwidth]{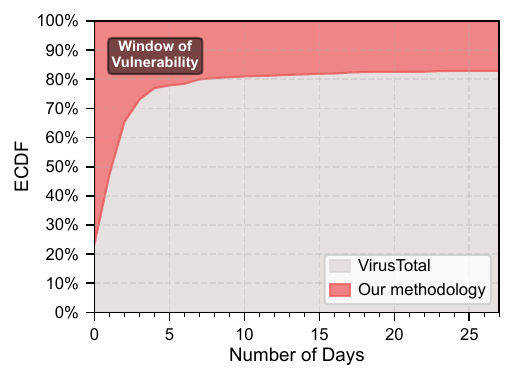}
        \caption{Distribution of number of days required for \vt to flag domains as malicious. Domains are detected by our method on day zero. The shaded area indicates the window of vulnerability during which domains remained undetected by \vt.}
        \label{fig:window-of-vulnerability}
    \end{minipage}
    \hfill
    \begin{minipage}[t]{0.32\textwidth}
        \centering
        \includegraphics[width=1\columnwidth]{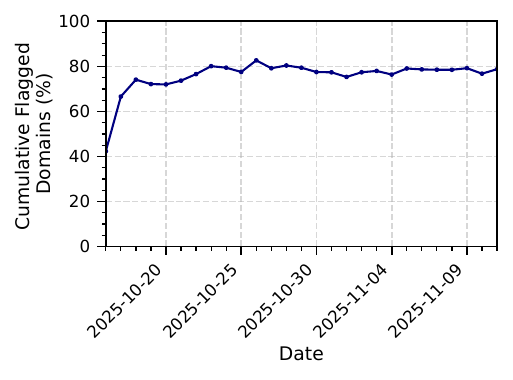}
        \caption{Longitudinal comparison of detected suspicious domains over a one-month period. \vt successfully detected $\sim$80\% of the domains discovered by \toolname, leaving a 20\% gap undetected throughout the study period.}
        \label{fig:longitudinal}
    \end{minipage}
\end{figure*}

To further understand these processes, we compute the ``registration rate'' of each website.
We define the registration rate of a website as the total number of domains in its cluster divided by the time interval (in seconds) between the first and last registration.
We plot in Figure~\ref{fig:registration_times} the registration rate for each domain in descending order with $y$-axis in log scale.
Each cluster contributes a number of points proportional to the number of domains it contains (\ie larger clusters have greater influence).
This emphasizes the registration patterns of large-scale domain campaigns.

We observe that there are high registration rates with an average rate of 1.3 domains per second, corresponding to more than 78 domains being registered every minute.
Additionally, the top 15\% of clusters exhibit extreme automated bursts with at least 120 domains registered per minute.
These rates exceed the rate at which a domain can be registered manually.
We also observe extreme cases with a maximum value of 83 domains registered per second.
This rate is observed for a cluster of 83 domains with the prefix \texttt{aqsmimg} followed by a number (\eg \texttt{aqsmimg1046.sbs}), all registered within the exact same second.
These patterns often indicate malicious infrastructure as legitimate users rarely need dozens of extremely similar domains at the exact same second.
Attackers register domains in bulk to rotate them quickly when some are blocked or taken down.

\finding{We demonstrate that \toolname uncovers instances of automated bulk registration.
The median cluster of detected domains is registered within just five minutes, with the average cluster containing domains registered at a rate of more than one domain every second.}

\section{\uppercase{Threat Exposure}}
\label{sec:findings}

\subsection{Window of Vulnerability}
\label{sec:window-of-vulnerability}

The novelty of our proposed methodology lies in that it shifts the defense against malicious domains to prevention rather than reaction.
Early detection of suspicious domains at registration time prevents harm to users and, more importantly, minimizes the attacker's operational window.
To measure the impact of our method, we measure the window of vulnerability the user is exposed to.
The window of vulnerability is the time period when a suspicious domain is registered until it is discovered by threat intelligence services.
During this time, users are exposed to potential harm.

We follow the data collection process of Section~\ref{sec:data-collection} and download newly registered domains of 6 different TLDs on a daily basis.
We apply our detection methodology on these domains and successfully identify suspicious domains that exhibit characteristics of bulk registration.
We run this experiment for one month starting October 2025, adding newly detected domains daily.
Each day, 20\% of these domains are randomly sampled and evaluated using \vt.
We evaluate both the domains that \toolname identifies each day, and those that were identified on prior days but remained unflagged by \vt.
We perform this analysis daily until a domain is flagged by \vt.
The sampling of domains is necessary to ensure that we can timely check all domains due to resource limits.
The experiment concludes after 27 days with more than 26K detected suspicious domains, of which more than 18K (69\%) are determined to be active at some point throughout the experiment.

In addition to this, similar to the process described in Section~\ref{sec:detection-validation}, we manually examine domains detected by \toolname to verify that they indeed exhibit suspicious, or even malicious, behavior.
Towards this, we examine URLs embedded on the website, redirections to other domains and smartphone applications available for download on websites.
Even though \vt might not label the domains themselves as suspicious, it acknowledges that they contain suspicious content or redirect users to malicious domains.

For each domain our methodology classifies as suspicious, we compute the absolute number of days it takes for \vt to detect it as suspicious or malicious.
We plot in Figure~\ref{fig:window-of-vulnerability} the cumulative distribution of the number of days it takes for \vt to flag domains.
\toolname classifies domains as suspicious at registration time (\ie day zero).
The red-shaded area therefore represents the window of vulnerability during which domains remain undetected by \vt security engines, exposing users to potential harm.
We observe that the median malicious domain is identified by \vt within two days.
Notably, 80\% of suspicious domains are identified within the first 7 days, indicating that the remaining top 20\% of suspicious domains take longer than a week to be detected or are not even detected within the observation period.
In fact, there is a steep rise in the left-hand side of Figure~\ref{fig:window-of-vulnerability}, indicating that a big portion of suspicious domains are detected within just a few days.
However, after day 5, there is a very slow rise and the plot reaches a plateau at approximately 83\%.
This slow rise corresponds to domains that take longer to be detected and the plateau indicates that there are almost no detections after that.

\finding{
Our method detects malicious domains immediately at registration, while \vt can take up to 7 days to flag them, leaving a substantial window of vulnerability.
}

Suspicious domain activity can be temporal since adversaries might register domains in bursts or change their naming patterns~\cite{sood2016taxonomy}.
Towards that extent, we also evaluate the performance of our detection methodology in a real-world setting, where new domains are registered daily and must be identified in a continuous, rolling manner.
We track whether and when each domain is detected by \vt and, for each day since registration, calculate the cumulative percentage of domains that have been detected up to that day.
This allows us to determine whether \vt classifies domains as suspicious given additional time, indicating delay in detection.

We plot in Figure~\ref{fig:longitudinal} the cumulative detection rate of suspicious domains by \vt.
For each day $d$, we plot the percentage of suspicious domains detected by \vt up to, and including, day $d$.
We observe that \vt reaches a plateau, and across the one-month period it detects $\sim$80\% of domains as suspicious or malicious.
The remaining 20\% are never detected by \vt at any point during the study window, even though a manual analysis using \vt's own engines, explicitly proves they are indeed suspicious.
Domains that \toolname flags as suspicious at registration time (\ie day zero), are not detected by \vt even though it has multiple days of additional observations to analyze.
This 20\% of remaining domains exceeds the false positive rate and can be attributed to websites of low activity, cloaked or short-lived domains~\cite{lim2025registration}.
Since \toolname identifies these domains at the moment of registration, it provides early and more complete awareness.
We acknowledge that \toolname focuses on a specific subset of threats and does not attempt to provide comprehensive coverage of all malicious domains.

\finding{
We discover that there is a persistent 20\% of domains missed by \vt compared to \toolname's zero-day detection.
\toolname extends on systematic blind spots of security engines and provides additional value for early warning.
}

\subsection{Content-Agnostic Detection}

A key feature of \toolname is its content-agnostic detection capability allowing us to identify suspicious domains without analyzing their content.
In Section~\ref{sec:window-of-vulnerability}, we compare our system against \vt, demonstrating it can successfully detect suspicious or, in some cases, malicious domains days in advance.
For a fair comparison, we ensure that \vt also provides a content-agnostic detection module to detect domains without examining their content.

In the longitudinal experiment of Section~\ref{sec:window-of-vulnerability}, we monitor the set of sampled domains on a daily basis to assess their activity status.
For this analysis, we only select domains that remain inactive throughout the entire observation period.
We define domains as inactive if they don't resolve to an accessible address or they respond with a status different from 2xx or 3xx.
We then compare the performance of \toolname with \vt on this set of inactive domains.

\begin{figure}
    \centering
    \includegraphics[width=0.8\columnwidth]{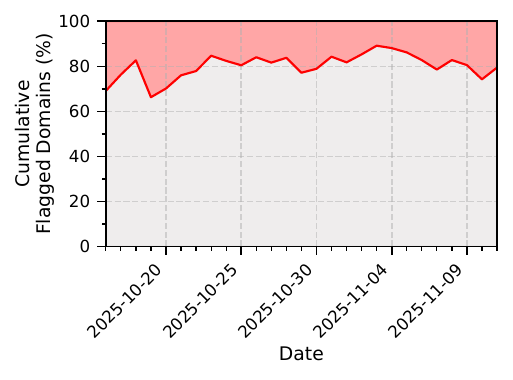}
    \caption{Cumulative proportion of inactive domains flagged as suspicious or malicious by \vt. 100\% represents the total set of domains identified by \toolname, illustrating the accumulated detected suspicious domains.}
    \label{fig:inactive-domains}
\end{figure}

We plot in Figure~\ref{fig:inactive-domains} the cumulative proportion of inactive domains flagged by \vt as suspicious over the period of the one-month observation window.
The $y$-axis shows the percentage of \emph{inactive} suspicious domains detected, with 100\% representing all domains identified by \toolname.
Each $x$-axis point represents a day, with the $y$-value showing the cumulative proportion of \emph{inactive} domains detected as suspicious by \vt up to that day.
We observe that \vt consistently identifies between 70\% and 90\% of the domains flagged by our system.
This indicates that while it detects a substantial portion of suspicious domains, it consistently misses 10–30\% of domains identified by \toolname.
The somewhat stable performance of \vt highlights that the increased coverage and earlier detection of \toolname's content-agnostic approach is persistent over time and not attributed to temporal inconsistencies.

Notably, our experimental results demonstrate that \vt's engines provide a content-agnostic detection functionality since it can discover suspicious domains that have never been active.
By establishing that \vt also employs a content-agnostic approach, we ensure that the evaluation on the efficacy and promptness of \toolname is valid rather than due to differences in the underlying detection methodology.

\finding{
We demonstrate that \vt also employs content-agnostic detection, showing that \toolname's ability to detect suspicious domains earlier and more complete comes from its design rather than differences in underlying detection strategies (content-only versus content-agnostic).
}

\subsection{Domain Category Evaluation}

Malicious domains used for different purposes can exhibit different behavior regarding their lifespan and traffic volume.
To investigate \toolname's performance across different types of websites, we conduct a category analysis.
Towards that, we perform a manual annotation task using independent reviewers to determine the category of domains.
We randomly sample 15\% of domains from the deployment described in Section~\ref{sec:window-of-vulnerability}, amounting to 2,825 distinct domains.
Using an automated crawler we capture screenshots of these domains from different time windows.

Next, websites are manually labeled into specific suspicious categories, building on the publicly available taxonomy from Cyren~\cite{cyrenCategoriesDescriptions}.
The authors of this work have no affiliation with this service.
We focus on Security, Parental Control and Productivity categories, grouping subcategories under their parent category, and adjusting them to reflect our collected data (\eg domains returning error codes like 403, 404, or 500 are labeled as ``Network Error'').
The annotation is performed by two independent reviewers, one of which is an author of this work, while the second is not affiliated with this work but has a computer science background. 
Reviewers have access to screenshots of the domains, as well as the definition of each category and are allowed to directly browse the domains or use any translation tool they deem necessary.
Finally, reviewers perform the manual annotation tasks separately and are not allowed to interact with each other during the experiment.
After manual annotation, reviewers agreed on the category of 2,550 domains, reflecting an agreement score of 90.3\% with a Cohen Kappa Score of $0.88$.
We retain only domains where both reviewers agree on the category, excluding all others to ensure high confidence.

\begin{figure}
    \centering
    \includegraphics[width=0.8\columnwidth]{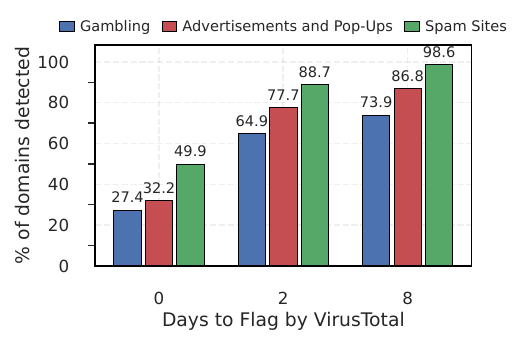}
    \caption{Distribution of number of days required for \vt to flag domains in different categories. \vt demonstrates variable detection speed, performing well for some categories but delaying in others.}
    \label{fig:categories_performance}
\end{figure}

We discover that the most prominent category in our sampled dataset is ``Network Errors'' with 725 distinct domains (28.4\%) that do not resolve to any valid IP address or respond with an error code.
Next, are ``Spam'' websites with 607 distinct domains (23.8\%) that contain websites that promote content through spam techniques.
``Gambling'' and ``Adult'' websites are the next most popular categories with 22.7\% and 7.7\% of total domains, respectively.
Finally, we observe small numbers of parked domains and websites that primarily serve advertisements.

Next, we study the distribution of number of days required for \vt to flag domains of each category as suspicious or malicious.
Our results show that while \toolname maintains consistently high detection performance across all categories, \vt exhibits variable accuracy, performing well in some categories but underperforming in others.
We plot in Figure~\ref{fig:categories_performance} the percentage of ``Gambling'', ``Advertisements'' and ``Spam'' domains identified by \vt on day 0, day 2 and day 8.
All domains of each category have been identified as suspicious at registration time by \toolname, corresponding to the 100\% ceiling of the figure.
We observe that on day zero only 27\% of suspicious gambling domains are identified compared to nearly half of spam websites.
Two days after their registration, \vt has successfully identified the majority of suspicious or malicious domains across categories.
\toolname performs consistently across multiple domain categories, while \vt's focus seems to be specialized toward specific types of threats.
In fact, previous work has discovered that the security scanners of \vt specialize in different attack types, with no single scanner performing well for all types~\cite{choo2023large}.
\vt's best performance is observed for Spam websites, with 98.8\% of domains detected and the median website being detected the date it is launched.
However, for suspicious Gambling websites, it reaches a plateau at around 80\%, failing to identify 20.1\% of suspicious or malicious gambling websites.
Although our system works well across different categories of threats, it is designed to work alongside established platforms like \vt, which continue to provide robust coverage across a wide range of malware.

\toolname is content-agnostic by design, and therefore does not directly rely on website category information. 
Figure~\ref{fig:categories_performance} empirically demonstrates that the method's performance remains stable across domains that host different types of content.
Although clustering operates purely at the lexical level, it is important to verify its effectiveness is not biased toward domain names associated with specific categories.

\finding{
Our results show that \toolname is robust across diverse website types and not dependent on category-specific characteristics, enabling deployment across a range of suspicious domains.
}
\section{\uppercase{Real-World Deployment}}
\label{sec:real-world-deployment}

To evaluate the real-world applicability of our methodology, we deploy \toolname over three distinct time windows between September and November 2025.
This allows us to assess its performance under varying temporal conditions and domain populations.
During each deployment, newly registered domains for the 6 different TLDs are continuously analyzed at registration time (\ie day zero), and any domains flagged as suspicious are subsequently compared against \vt for external verification. 

Across the full deployment period, we process $1.5$M distinct domains, with \toolname successfully identifying 47,345 domains as suspicious on day zero, demonstrating its capability to operate at scale in diverse but practical environments.
To support further research, we make the full set of detected suspicious domains publicly available~\cite{openSource} and discuss ethical considerations in Section~\ref{sec:conclusionDiscussion}.
Over 40K of them (85.36\%) are also classified as suspicious or malicious by \vt at some point in time.
Furthermore, for each active domain in our dataset, we also examine embedded URLs, network redirections and downloaded applications and feed them to \vt for further security analysis.
This analysis allows us to validate the performance of \toolname as it discovers domains that \vt initially misses but are later confirmed malicious.
Altogether, 44,858 of them (94.75\%) are suspicious or malicious either directly or due to the content they distribute.
Interestingly, 65\% of the domains \toolname flagged are in fact labeled malicious, and 35\% are suspicious.

These results underscore the practical effectiveness of our approach.
Detecting such a large number of malicious domains immediately at registration demonstrates that our methodology not only generalizes to real-world conditions but also offers a substantial improvement in responsiveness compared to conventional methods.
Early detection of these domains significantly reduces the window of vulnerability.

\finding{
Using three months of real-world domain registration data, \toolname detects 47,345 malicious or suspicious domains the day they are registered.
In contrast, \vt verifies 94.75\% of these domains only after a delay, highlighting \toolname's effectiveness in early detection.
}

We further analyze detected malicious domains to understand their operation and the strategies used by bad actors.
First, we study their popularity using the Tranco ranking~\cite{tranco}.
From 2025-09-10 to 2025-11-25, we examine whether any of the detected suspicious domains appear in the daily Tranco list, evaluating them against 6.3M distinct websites.
We discover that the majority of the detected domains do not appear in Tranco rankings, with only 612 distinct domains being ranked at least once during the observation period.
This is expected as these websites have a short lifespan of a few days.
Their low visibility and limited legitimate use keeps them from being popular.
The highest ranked domain is at position 102K indicating that detected domains are unlikely to attract substantial legitimate traffic.

\begin{table}[t]
\centering
\scriptsize
\begin{tabular}{lrr}
\toprule
\textbf{Registrar} & \textbf{\# suspicious} & \textbf{\% suspicious} \\
& \textbf{domains} & \textbf{domains} \\
\midrule
GoDaddy.com, LLC & 20,858 & 53.55\% \\
Gname.com Pte. Ltd. & 5,380 & 13.81\% \\
Key-Systems GmbH & 2,514 & 6.45\% \\
NameSilo, LLC & 1,776 & 4.56\% \\
Domain International Services Ltd & 1,762 & 4.52\% \\
\bottomrule
\end{tabular}
\caption{Top registrars responsible for the majority of registered suspicious domains. Registrars of suspicious domains are extracted using RDAP information for 38,951 domains.}
\label{tab:suspicious-registrars}
\end{table}

Next, we examine the registrars of suspicious and malicious domains.
Using the RDAP protocol, we successfully collect registration information for 38.9K (82.27\%) domains detected by \toolname.
In Table~\ref{tab:suspicious-registrars}, we observe that GoDadddy accounts for the majority of domains (53\%), registering 20.8K distinct suspicious or malicious domains.
Gname follows with 14\%, and Key-Systems with 6\% of suspicious domains, while all remaining registrars individually account for less than 5\% each.
Evidently, a small number of registrars are disproportionately used for the bulk registration of suspicious domains.
That is, specific registrars are preferred by suspicious actors conducting bulk domain registrations.
GoDaddy has already been found to be the registrar with the most malware domains~\cite{aaron2021malware}.

To contextualize these findings, we conduct a control experiment on 1,100 benign domains from the Tranco list, restricted to the same TLDs as our primary dataset and verified as benign via \vt.
In contrast to the suspicious and malicious domain set, the registrar distribution for benign domains differs substantially.
NameCheap accounts for 23.3\%, NameSilo for 17.2\%, Alibaba for 8.2\% and GoDaddy for 7.6\%.
Notably, registrars that dominate the suspicious domain ecosystem do not have a comparable share for benign domains, while those prevalent in the benign set account for only a minor fraction of suspicious registrations.
This discrepancy suggests that malicious actors disproportionately rely on a subset of registrars, potentially due to policy factors.

Finally, we study independent clusters of suspicious domains, which group domain names with high structural and lexical similarity.
We find that 87.3\% of these clusters contain websites under the same registrar, suggesting coordinated and bulk registration by the same actor.
We also identify 736 distinct clusters with domains under different registrars.
A manual analysis of these heterogeneous clusters reveals that 480 clusters use a combination of GoDaddy and ``Domain International Services Limited'', while 103 clusters use a combination of GoDaddy and ``Gname''.
Despite the different providers, domains in these clusters use common naming format with 3 digits followed by 2 letters (\eg \emph{855dj.top}).
We also observe patterns in timing activity with registration peaks at 14:00 for GoDaddy and shifts to ``Domain International Services Limited'' at 15:00 and Gname at 18:00 (under a single timezone).
We attribute these patterns to attempts to reduce detection risk across registrars.

\finding{
Suspicious domains are generally unpopular and disproportionally registered through specific registrars (most prominently GoDaddy).
While most clusters share a single registrar, others diversify, indicating attempts to avoid detection.
}
\section{\uppercase{Related Work}}

There is extensive prior work on malicious domain and URL detection, including lexical, machine learning and temporal analysis.
Lexical and URL-based methods have leveraged features such as URL tokens, N-grams, entropy and edit-distance similarity to detect phishing and algorithmically generated domains~\cite{le2011phishdef,ma2009beyond,selvi2019detection,zhang2017domain,zhao2019malicious}.
These approaches are effective for lightweight classification but often focus on active URLs and may require careful feature engineering.
Machine learning-based approaches have incorporated lexical, semantic, and brand-aware features using classifiers from logistic regression to deep NLP models~\cite{almomani2022phishing,liu2024less,shirazi2018kn0w,ccolhak2024securereg}.
While these methods improve detection accuracy, they typically require continual retraining and incur operational overhead.
DNS and temporal analyses exploit domain registration patterns and public records, and fast-flux behavior to identify suspicious domains~\cite{chiba2016domainprofiler,kusumi2025malicious,alsabah2022content,lim2025registration,lee20257}.
These approaches can prioritize manual investigation and reveal bulk registration strategies, but often rely on active campaigns or large-scale DNS data, limiting proactive detection.
Instead, \toolname identifies malicious domains at registration, without relying on content analysis, DNS traffic, or retraining.
By leveraging structural and lexical similarity, it enables early detection of bulk registrations before abuse, addressing gaps in traditional lexical, ML, and temporal methods~\cite{desmet2021premadoma}.

PREDATOR~\cite{hao2016predator} detects newly registered domains using 22 features, including edit distances to known malicious domains, digit ratios, and dormancy periods, relying on real-time zone files and historical blacklists.
On labeled data, with the false positive rate constrained below 0.35\%, it achieves a detection rate of 70\% and estimated precision above 80.95\%, although the exact value was not reported.
In contrast, our method targets a specific class of suspicious domains and achieves substantially higher reliability, with a precision of 98.5\% and a false discovery rate below 1.5\%.
PREDATOR offers broad threat coverage but with lower detection certainty, whereas our method achieves highly reliable detection within a narrower, practically important scope. 
The two approaches are therefore complementary, with our system particularly suited to settings requiring minimal false positives and high-confidence alerts.
 
A more recent approach leverages bulk-registration patterns to flag suspicious domains using clustering over registrar–nameserver combinations~\cite{cheadle2025detecting}.
While highly precise for large, bursty registration batches, the method cannot detect small and asynchronous registrations or link domains registered across multiple registrars.
Finally, MANTIS~\cite{deniz2025mantis} detects zero-day malicious domains by monitoring low-reputation hosting infrastructure and building a Graph Neural Network over Passive DNS and \vt.
While achieving high precision and volume, its effectiveness depends on ``toxic'' hosting neighborhoods, thus missing domains hosted on legitimate infrastructure.
In contrast, \toolname detects malicious domains at registration, without relying on historical blacklists, infrastructure assumptions, or complex graph models.
This enables scalable early detection of bulk-registrations.
\section{\uppercase{Discussion \& Conclusion}}
\label{sec:conclusionDiscussion}

\point{Summary:}
The proliferation of malicious and suspicious domains is a growing threat to online security, and early detection can substantially reduce exposure and damage.
In this work, we propose \toolname, a content-agnostic framework that identifies malicious domains at registration time, without relying on hosted content or user reports.
\toolname detects patterns of suspicious bulk registration capturing lexical and structural similarity.
We manually evaluate discovered domains and, through cross-verification with state-of-the-art security engines, we demonstrate that \toolname achieves 98.5\% precision.
Longitudinal analysis reveals that flagging domains at registration can reduce the exposure window by up to 7 days compared to popular threat intelligence engines.
In addition to this, we demonstrate that \toolname performs consistently across diverse website categories, detecting 20\% of suspicious domains missed by existing engines.
When applied to 1.5M real-world domains, it identifies 45K suspicious or malicious domains.

\point{Discussion:}
Traditional detection approaches rely on website content or hosting behavior, allowing adversaries to begin their campaigns.
\toolname focuses on prevention rather than reaction.
Detecting domains at registration enables proactive intervention before malicious activity occurs.
For coordinated actors, the interval between registration and activation can be extremely short, meaning campaigns may start almost immediately.
Flagging or suspending suspicious domains at registration can disrupt attacks, minimizing the operational window.
As shown in this work, registration patterns and timing can reveal suspicious bulk registrations, allowing defenders to identify entire campaigns rather than only individual domains. 

\point{Limitations:}
Newly registered domains are acquired with an up to 2-day delay due to the commercial provider we utilize.
This does not affect our findings since reported results are a lower bound on \toolname's performance and any evaluation delay underestimates its effectiveness.
Moreover, we acknowledge that a sufficiently adaptive adversary could evade \toolname by abandoning templated bulk registrations or introducing lexical diversity. 
However, this requires increased operational complexity, as automated large-scale campaigns rely on systematic naming for efficiency.
Our method is designed to detect cost-effective abuse patterns rather than all possible adversarial strategies.
It is intended as a complement, rather than replacement, to established services, which remain effective for broad malicious domain detection.

\point{Ethical Considerations:}
This work analyzes domain registrations using passive data from a public provider.
We do not interact with registrars or registrants.
For verification, we capture screenshots of publicly accessible landing pages once per day after full load, without performing any security scanning or probing.
The domains we release are associated with suspicious activity and are provided for research, educational, and informational purposes only.
Detected domains are cross-referenced with \vt, through which any malicious files (\eg Android apps) are submitted to the broader threat detection community.
Finally, in accordance with GDPR, we do not collect or process personal data.

\section*{\uppercase{Acknowledgments}}

Funded by the European Union (C-SOC grant id 8515915).
Views and opinions expressed are however those of the author(s) only and do not necessarily reflect those of the European Union or the European Cybersecurity Competence Centre.
Neither the European Union nor the European Cybersecurity Competence Centre can be held responsible for them.

\bibliographystyle{apalike}
\bibliography{main}

\end{document}